\documentstyle[latex8]{article}

\newtheorem{theo}{Theorem}

\newcommand {\bbox} {\vrule height7pt width4pt depth1pt}

\newtheorem{lemm}{Lemma}

\newtheorem{deff}{Definition}

\begin{document}
\title{Fault-Tolerant Quantum Computation With Constant Error}
\author{D.~Aharonov\thanks{
Institutes of Physics and Computer science, The Hebrew University,
Jerusalem, Israel, E-mail: doria@cs.huji.ac.il}
 \and M.~Ben-Or\thanks{
Institute of Computer science,
The Hebrew University, Jerusalem,
Israel, E-mail: benor@cs.huji.ac.il}}

\maketitle

\begin{abstract}
In the past year many developments had taken place in the area of
quantum error corrections.
Recently Shor showed how to perform fault tolerant quantum computation
 when, $\eta$, the probability for a fault in one time step per qubit
or per gate, is logarithmically small.
This paper improves this bound and shows
 how to perform fault tolerant quantum computation
 when the error probability, $\eta$, is smaller than some constant threshold,
$\eta_0$.
The cost is polylogarithmic in time and space,  
and no measurements are used during the quantum computation.
The same result is shown also for quantum circuits which operate
on nearest neighbors only.

To achieve this noise resistance, we 
use concatenated quantum error correcting codes.
The scheme presented is general, and works with any quantum code,
conditioned that the code satisfies some restrictions, namely that it is 
a ``proper quantum code''.
The constant threshold $\eta_0$ is a function of the parameters 
of the specific  proper code used.

We present two explicit classes of proper quantum codes.
The first example of proper quantum codes generalizes 
classical secret sharing with polynomials.
The second uses a known class of quantum codes and converts it 
to a proper code.  This class is defined over a field with $p$ elements,
which means that  the elementary quantum  particle is not a qubit but 
a ``qupit''.

We estimate the threshold $\eta_0$ to be $\simeq 10^{-6}$. 
Hopefully, this paper motivates a search 
for proper quantum codes with higher thresholds,
at which point quantum computation becomes practical. 
\end{abstract}

\section{Introduction}
Quantum computation is believed to be more powerful
than classical computation, mainly due to the celebrated 
Shor algorithm\cite{Shor94}.
It is yet unclear whether and how quantum computers will
 be physically realizable,\cite{Loyd93,DiV95,CZ}
but as any physical system, they {\em in principle}
will be subjected to noise, such as decoherence\cite{Zur91,Unr94,PSE95},
and inaccuracies.
Without error corrections, the effect of noise can 
ruin the entire computation\cite{Unr94,Chu95}, so
we need to protect the computation against quantum noise.
Already the question of protecting  quantum information is harder 
than the  classical analog because one should also protect the
 quantum correlations between the quantum bits (qubits).
However,  it was shown \cite{CS95,Ste}
 that good quantum error correcting codes 
exist, a result which was followed by many explicit examples(ex: 
\cite{Sho95,LMPZ}).
This does not immediately imply the existence of noise resistant 
  quantum computation, since  the computation causes  the effect 
of faults to spread.
 Recently Shor\cite{Sho96}
 showed how to use quantum codes in order to perform 
fault tolerant quantum computation when the {\it noise rate}, or the  
 fault probability each time step, per qubit or gate,
 is polylogarithmically small.
In this paper,  we close the gap  
and show how to perform fault tolerant quantum 
 computation in the presence of constant noise rate, with polylogarithmic
 cost  in time and size.

The error corrections which where described so far used a combination
of classical and quantum operations. 
We would like to define a model of noisy quantum computation\cite{AB96},  
such that errors and error corrections can be described 
entirely inside this model.
Working in the correct model will enable to prove the  result of this paper.
Sequential quantum computation can not be noise resistant\cite{AB96},
so we work with quantum circuits\cite{Deu89,Yao93}, and since 
the  state of  a noisy quantum system
is  in  general a
probability distribution over pure states, i.e. a {\it  mixed  state},  
 and not merely a pure state as in
the standard model, we use quantum circuits with mixed states\cite{AN96}.
Since noise  is a dynamic process which depends on time,
the circuit will be divided to levels, or time steps.
 Unlike Yao\cite{Yao93} we allow 
qubits to be input and output at different times to the circuit.
This is crucial, since with the restriction that all qubits
are initialized at time $0$, it is not possible to compute 
fault tolerantly without an exponential blowup in the size of
 the circuit\cite{ABIN}.
Between the time steps, we add the noise process, which is a probabilistic 
process:
 each qubit or gate  undergoes  a fault
 with independent probability $\eta$ per step,
 and $\eta$ is referred to as the {\it noise  rate}.
The list of times and places where faults had occurred, namely
 the {\it fault path},
 is random, and naturally, the function that the
 circuit computes is a weighted average
over the noise process.

{~}

\noindent{\bf Computing on encoded states:}
Let us first describe how one can protect quantum information against noise,
using quantum error correcting codes\cite{CS95,Ste}.  
The idea behind these codes is, as in classical linear block codes,
 to spread the state of one qubit 
on a number of qubits, called a ``block'', 
such that even if some of the qubits in a block are damaged,
the state of the system can be recovered from the other correct qubits.
If we want to preserve the state of $n$ qubits, we encode it on $n$
blocks. The whole state should be recoverable if 
not too many errors occurred in each block.
Quantum codes that can correct $d$ faults 
have the property that if not more than $d$ faults
occurred in each block, there is some operation that can recover 
the encoded state given the damaged state.
The difference from classical codes is that the quantum correlations should 
be recovered as well, and not only the logical states of the qubits.
However, it turns out that applying  classical error corrections in one
 basis of the Quantum space can correct the logical states, while applying
classical error corrections in some other basis corresponds to corrections 
of the quantum correlations. 

In order to protect a quantum computation against faults, one 
can try to compute on encoded states, and not on the states themselves,
using quantum codes.
The circuit $M_1$ which will compute on encoded states
is defined as a simulation of the original circuit $M_0$.
A qubit in the original circuit transforms  to a block of qubits
 in the simulating circuit,
and each time step transforms
  to a {\it working period} of several time steps.
To simulate the computation done in the $t'th$ time step in $M_0$,
 we will apply in $M_1$ the analog computation, on encoded states.
Each gate in $M_0$ will transform to some ``procedure'' in $M_1$,
which computes this gate on encoded states.
The procedure might require ancilla 
qubits, so these are added to the circuit $M_1$, and are initialized only
when needed.
If $M_1$ is initialized with some encoded state,
 we expect this state to evolve by the computation
along some ``trajectory'', such that at the end of each working period
it encodes the correct corresponding state of $M_0$.
The input and output to $M_1$ will simply  be a duplication of the 
inputs and outputs of $M_0$, on the corresponding blocks.
We will therefore need to add on each block, before 
computation begins, an {\it initialization procedure},
 that transforms the duplicated input, i.e a
 string of  $0's$ or a string of $1'$s, to the encoding state $|S_0>$,
or $|S_1>$.
At the end of the computation  we will need the opposite transformation,
so we will use a   {\it reading Procedure} that transforms a block in the 
state 
 $|S_0>$ or $|S_1>$ back  to a string
of $0$'s or $1'$s.

Computing on encoded quantum states does not
automatically provide protection against faults, since 
errors accumulate, and 
when the damage has effected  too many qubits in one block,
the correct state is no longer recoverable.
In order to be able to correct the state,
 error corrections should be applied all the time,
so that the error can not accumulate. 
Therefore in each working period in $M_1$ we will add a step of error 
corrections of each block.
The working period will be divided to two stages: a computation stage
and a correction stage.
The idea is therefore to apply alternately computation stages and
 correction stages, hoping that during the computation stage 
the damage that accumulated is still small enough so that 
the corrections are still able to correct it.
One should notice that this  ``hope''  not always comes true.
During the
computation faults can ``spread'' to places not in the fault path.
This spread can happen if a gate operates on a damaged qubit and some 
``correct'' qubits- in general, this can cause all the qubits that 
participate in the gate to be damaged.
If for example, a gate procedure consists of one gate operating on the
 whole block, than one fault is already unrecoverable.
The procedures should be designed in such a way that 
 a fault  can not effect all the qubits in the block.
In general, a fault at time $t$ in qubit $q$ can effect  
qubit $q'$ at time $t'>t$ if  there is a path in the circuit
 connecting the points $(q,t)$ and $(q',t')$.
Since we want to be able to correct after each computation stage,
We can define the ``spread'' of a procedure as the maximal number of qubits
in each block in the output of the procedure, which are effected by
a fault that happened in this procedure.
If we use only procedures with small spread,
then if not too many errors happened during the computation stage
 in each procedure, the error corrections will still be able  to 
 correct the damage using the other undamaged qubits.
We are actually looking for a pair of a
 quantum code which can correct $d$ errors, 
and a corresponding set of {\it universal} gates, such that
their procedures, with respect to the code,
 allow one fault to spread to at most $d$ qubits.
Since the error corrections, initialization, and reading procedures
 are also subjected to faults, we need them to
have small spread too. 
Such  codes, with the corresponding universal set of gates,
 will be called {\it quantum computation codes}.

We now want to show that the reliability of the simulating circuit
 is larger than the original circuit. 
In the original circuit, if one fault occurred the  
computation failed, but the simulating circuit can tolerate
a number of errors, say $k$, in each procedure, since the error corrections 
correct them afterwards.
The effective noise rate of $M_1$
 is thus the probability for more than $k$ errors in
a procedure, and it will be  smaller than the actual noise rate $\eta$, 
if the parameters are chosen correctly.
However, it seems that an improvement from a constant noise rate
to polynomially small effective noise rate, as we need for fault tolerance,
is hard to get in the above scheme.
In \cite{Sho96} it is shown how to apply the above scheme, with a specific 
quantum computation code, to achieve 
fault tolerance when the noise rate $\eta$ is taken to be
 logarithmically small.


{~}

\noindent{\bf Improvement to constant noise rate}
To improve this result, we use {\it concatenated simulations}, 
which generalizes the works of Gacs \cite{Gacs89} to the quantum case.
The idea 
is that since the simulating circuit is
 also a circuit, it's effective noise rate
can be improved by simulating it again, 
and so on
 for several levels. 
Even a very small improvement in each level will suffice since the 
improvement is exponential in the number of levels. 
Such a small improvement can be achieved when using a code of constant 
block size, if the noise rate is smaller than some threshold $\eta_0$.
Note, that each level simulates the error corrections in the 
simulated level, and adds error corrections in the current level.
The final circuit thus includes error corrections of all the levels,
where the computation of error corrections of high levels is corrected all 
the time in small levels.
The lower the level, the more often are
error corrections of this level applied. This is in correspondence with 
the fact that lower levels work on smaller blocks,
 that are more likely to be quickly damaged.
The whole scheme relies heavily on the fact that the places where
 errors occur are random and independent.
Most of the faults will probably be corrected by error corrections
in the first level,
and if this does not work, than using other blocks which where 
corrected in the first level, the error will  probably be corrected 
in the second level. The probability that the highest blocks 
can not be recoverable is polynomially small.

Not any quantum computation code can be used in the above scheme.
There are two restrictions:(1)
When applying the simulation, we replace the gates by fault tolerant
procedures.
Since we want to simulate the new circuit as well, we need
that these procedures use gates that can be replaced by 
fault tolerant procedures as well.
Hence the universal set of gates associated with the
 quantum computation code that we use, must have fault tolerant
procedures which use gates from the same universal set of gates.
This is the ``closeness'' restriction.
(2) Let us consider a two level simulation.
If a simulated error correction operates on an encoded wrong word, it clearly
corrects it.
But what happens if it gets as an input some state which 
does not encode any word?
The simulated error correction ``understands'' only encoded words.
If we demand that error correction in the lower level
 corrects any state to some word in the code, than the input for the  
error correction of the upper level will maybe be 
 wrong but it will be
 an encoded word, so it can be understood and corrected by the upper level. 
The second restriction is therefore that the error correction 
takes any state to some word in the code. 
Quantum computation codes which satisfy both restrictions are called 
{\it proper quantum computation codes}.

{~}

\noindent{\bf Explicit proper codes over $F_p$}
We describe  two classes of proper quantum computation codes.
We first describe the class of quantum codes in \cite{CS95} 
generalized to codes over $F_p$, for any prime $p$.
These codes are defined for general  quantum  circuits which 
 consist of particles with $ p\ge 2$ possible states to be in.
 We call such quantum particles
{\it qupits}, as a generalization to qubits.
Non binary codes where defined independently also 
by Chuang\cite{Chu96} and Knill\cite{Knill}.
The proofs that these are quantum codes\cite{CS95}
 transform smoothly  from $F_2$ to $F_p$.
For $p=2$, we show how to convert these codes to proper quantum codes, using 
the set of gates and procedures described in \cite{Sho96},
which are modified to fit the definition of 
proper quantum codes.
The second example of proper quantum codes 
uses a special case of the linear quantum codes,
which is the quantum analog of random polynomial codes\cite{BGW}. 
The requirements imposed on quantum computation codes are actually
very similar
to the requirements that are imposed on the secret sharing schemes
that are used to perform secure fault-tolerant distributed computation
\cite{BGW}. In both cases measuring a few qubits, or reading a few
secret shares, must give no information about the unencoded data,
and in both cases we must show how to perform computation
in a fault-tolerant way on the encoded data. 
To adopt the techniques of \cite{BGW} to the quantum setting one
can use the same encoding but instead of selecting a random
polynomial to share a secret we simply pick the superposition
of all those polynomials.

The noise rate threshold is shown to be larger  than $\simeq 10^{-6}$.  

{~}

The results of this paper hold also with
 a more general  noise model, in which  
 for all integer $k$, for any set of $k$ 
points, the probability that a fault occured in all the points is bounded by $\eta^k$,  
and  an adversary can pick a 
transformation that operates on all the damaged qubits together.
The same result is true for quantum circuits which are
 allowed to operate only on nearest neighbor qubits
 (In this case the threshold will be smaller.)

Similar results to those of this paper
 where independently discovered by Knill, Laflamme and Zurek\cite{KLZ}.


%

{\bf Organization of paper:} In section 2 we define the model of noisy 
quantum circuits. 
Section 3 is devoted to describe one step of the simulation,
given any  quantum computation code.
In section 4 we present concatenated simulations and prove that
this scheme provides noise resistant computation  in the presence of
 a constant noise rate, given any proper quantum code. 
In section 5 we present two explicit classes of
 proper quantum computation code.
Section 6 discusses possible extensions, conclusions and open questions.

\section{Noisy Quantum Circuits}

In this section we
 recall the definitions of quantum circuits\cite{Saq,Deu89,Yao93}
 with mixed states\cite{AN96},
We then define noisy quantum circuits, which model
 a physical realization of quantum 
circuits\cite{AB96}.
The faults are probabilistic, and occur in single qubits and in gates.

\subsubsection{Pure states} 
 We deal with systems of  $n$ two-state quantum
particles, or ``qubits''. The {\em pure state} of such a system 
is a unit vector, denoted $|\alpha\rangle$,
in the $2^{n}$ dimensional complex space $\cal{C}$$^{2^{n}}$.
We view  $\cal{C}$$^{2^{n}}$ as a
 tensor product of $n$ two dimensional spaces, each corresponding to a qubit:
$\cal{C}$$^{2^{n}}= \cal{C}$$^{2}\otimes...\otimes\cal{C}$$^{2}$.
As a basis for   $\cal{C}$$^{{2}^{n}}$,
we use the $2^{n}$ orthogonal {\it basic states}:
 $|i\rangle=|i_{1}\rangle\otimes
|i_{2}\rangle....\otimes|i_{n}\rangle,0\le i< 2^{n}$,
 where $i$ is in binary representation,
and each $i_{j}$ gets 0 or 1.
A general unit vector $|\alpha\rangle$ in  $\cal{C}$$^{2^{n}}$ is
 called a ``pure state'', and    
is a {\em superposition}
 of the basic states:  
$|\alpha\rangle = \sum_{i=1}^{2^{n}} c_{i}|i\rangle$, 
 with $\sum_{i=1}^{2^{n}} |c_{i}|^{2}=1$. $|\alpha\rangle$ corresponds
to the vector 
 $v_{\alpha}=(c_{1},c_{2},...,c_{2^{n}})$.
 $v_{\alpha}^{\dagger}$, the complex conjugate of  $v_{\alpha}$,
is denoted  $\langle\alpha|$.
The inner product between $|\alpha\rangle$ and $|\beta\rangle$
is $\langle\alpha|\beta\rangle=
 (v_{\alpha},v^{\dagger}_{\beta})$.
The matrix  $v_{\alpha}^{\dagger}v_{\beta}$
 is denoted as  $|\alpha\rangle\langle\beta|$.
An isolated system of n qubits 
develops in time by a unitary matrix,
 of size $2^{n} \times 2^{n}$:
 \( |\alpha(t_{2})\rangle = U|\alpha(t_{1})\rangle.\)
A quantum system in $\cal{C}$$^{{2}^{n}}$ can be {\em observed} by 
{\em measuring} the system.
An  important
 measurement is a {\it basic 
measurement} of
 a qubit $q$, of which the possible outcomes are $0,1$.
For the state  $|\alpha\rangle=\sum_{i=1}^{2^{n}} c_{i}|i\rangle$,
the  probability for outcome $0$ is $p_{0}= \sum_{i, i|_{q}=0}|c_{i}|^{2} $
and the state of the system 
will {\em collapse} to 
$|\beta\rangle=\frac{1}{p_{0}}\sum_{i, i|_{q}=0} c_{i}|i\rangle$,
 (the same for $1$).
A unitary operation $U$ on $k$ qubits  
 can be applied  on n qubits,
$n\geq k$, by taking the {\bf extension} $\tilde{U}$  of $U$,
i.e. the tensor product of $U$ with  an identity matrix on
 the other qubits. 
The deffinition can be generalised to circuits which operate on $p-$state
 quantum particles, or {\it qupits}. (simply replace
$2$ by  $p$ in the
definitions above).

\subsubsection{ Mixed states}
 A system which is not ideally isolated  from 
it's environment is described by a {\em mixed state}.
There are two equivalent descriptions of mixed states:
mixtures and density matrices.
We use density matrices in this paper.
A system in the {\bf mixture}  $\{\alpha\}=\{p_{k},|\alpha_{k}\rangle\}$
is  with probability $p_{k}$
in the pure state $|\alpha_{k}\rangle$.
The rules of development in time and measurements 
for mixtures are 
obtained by applying {\bf classical} probability to
the rules 
for pure states.
A density matrix $\rho$ 
on $\cal{C}$$^{2^{n}}$ is an hermitian positive semi definite complex matrix
 of dimensions $2^{n}\times 2^{n}$,
with $tr(\rho)=1$.
A pure state $|\alpha\rangle=\sum_{i} c_{i}|i\rangle$ 
is associated the density matrix
 \(\rho_{|\alpha\rangle} = |\alpha\rangle\langle\alpha|\) i.e.  
\(\rho_{|\alpha\rangle}(i,j)= c_{i}c_{j}^{*}.\)
A mixture 
$\{\alpha\}=\{p_{l},|\alpha_{l}\rangle\}$,
is  associated the density matrix :
\(\rho_{\{\alpha\}} = \sum_{l} p_{l} \rho_{|\alpha_{l}\rangle}.\)
The operations on a density
matrix are defined such that  the correspondence to mixtures is preserved.
If a unitary matrix $U$ transforms the mixture 
\(\{\alpha\}=\{p_{l},|\alpha_{l}\rangle\}\) to
\(\{\beta\}=\{p_{l},U|\alpha_{l}\rangle\},\)
then
 \(\rho_{\{\beta\}} = \sum_{l}  p_{l}
 U|\alpha_{l}\rangle\langle\alpha_{l}|U^{\dagger}=
U\rho_{\{\alpha\}}U^{\dagger}.\)
A basic measurement of the $j'$th qubit in $\rho$
gives, the outcome $0$ with the probability which is
 the sum of the diagonal terms of $\rho$, which relate to 
the basic states $i$ with $i_j=0$:
$pr(0)=\sum_{i=1}^{2^{n}} \rho_{i,i} \delta(i_j=0)$.
conditioned that the outcome is the eigenvalue $0$,
the resulting density matrix is $O_{0}\circ(\rho)$, which is the minor 
of $\rho$ which includes only  rows and columns which relate 
to basic states $i$ with $i_j=0$.
(This minor should of course be normalized to have trace one).
Without conditioning  on the outcome 
the resulting density matrix will be  
\(O\circ(\rho)=
 Pr(0) O_{0}\circ(\rho)+Pr(1) O_{1}\circ(\rho). \)
which differs from  $\rho$, only in
 that the entries in  $\rho$ which connected between
 $0$ and $1$ on the same qubit are put to zero.
Given a density matrix $\rho$ of n qubits,
 the reduced density matrix of a subsystem,$ A$,
 of, say, $m$ qubits is defined as an average over the states of
the other qubits:   
 \( \rho|_{A}(i,j)= \sum_{k=1}^{2^{n-m}} \rho(ik,jk)\).

\subsection{\bf Quantum circuits with mixed states}
{\em A quantum unitary gate}, $g$, of order $k$ is a complex unitary 
matrix of size $2^{k} \times 2^{k}$. 
A density matrix $\rho$ will transform
by the gate to  \(g\circ\rho = \tilde{U}\rho\tilde{U}^{\dagger}\),
where $\tilde{U}$ is the extension of $U$.
{\em A Quantum circuit} 
 is a directed acyclic graph
with $n$ inputs and $n$ outputs. 
 Each node $v$ in the graph is labeled by a quantum gate $g_{v}$.
The in-degree and out-degree of $v$ are equal to the order of $g_{v}$.
Some of the outputs are labeled ``result'' to indicate that
these are the qubits that will give the output of the circuit.
The wires in the circuit correspond to qubits.
An initial density matrix $\rho$ 
transforms by a circuit $Q$ to a 
 final density matrix  $Q\circ \rho =
g_{t}\circ...\circ g_{2}\circ g_{1}\circ \rho$,
where the gates  $ g_{t}...g_{1}$ are applied in a topological order. 
For an input string $i$,
the initial density matrix is $\rho_{|i\rangle}$.
The output of the circuit is the outcome of applying basic measurements
of the result qubits, on the final density matrix  
$Q\circ\rho_{|i\rangle}$. Since the outcomes of measurements
 are random, the function that the circuit computes is a 
{\it probabilistic function}, i.e. for input $i$ it outputs 
strings according to a distribution which depends on $i$.

\subsection{Noisy Quantum Circuits}
As any physical system, a quantum system is subjected to noise.
The process of noise is dynamic and depends on time.
We therefore divide the   quantum circuit to levels, or time steps.
We permit that qubits are input and output at different
times, and we say a qubit is {\it alive}
from $t_{1}$ to $t_{2}$ if it is input to the circuit 
at   $t_{1}$ and output at $t_{2}$.
The ``space time'' of the noisy quantum circuit is a two dimentional array,
 consisting of all 
 the pairs $(q,t)$, 
of a qubit $q$ and time $t$, where the qubit $q$ is alive at time $t$.
The volume of the circuit $M$,  is the number of points in it's
 space time and it is denoted by $V(M)$.

In our model of noisy quantum circuits
 each qubit and each gate are 
 damaged with probability $1/2>\eta>0$ each time step.
The damage operates as follows: 
A unitary operation operates on the qubit, (or on the qubits that 
are output from the gate in the case of a gate damage)
 {\it and} on a state of
 the environment (The environment can be represented by $m$
 qubits in some state).
This operation results in a density matrix of the $n+m$ qubits.
 We reduce this density matrix to the $n$ qubits of the 
circuit to get the new density matrix after the damage.
A noise step happens between two levels, or two time steps of computation,
and in a noise step each gate or qubit is damaged with probability $\eta$.
The density matrix of the circuit develops by applying alternately 
computation steps and  noise steps.
Each ``run'' of the computation is subjected to a specific ``fault path'',
which indicates where and when fault occured.
Each run ends up with some output.
The function computed by the noisy quantum circuit is naturally 
the average over the outputs, on  the probabilistic process of noise.
%

\section{Quantum computation on encoded states}

In the following section  we   define  quantum codes
and quantum computation codes.
Then we describe how to improve reliability of computation using 
quantum computation codes.
Quantum computation codes are used  
abstractly in this section, while
explicit examples of such codes are given in section \ref{explicit}.

\subsection{Quantum block codes}

A quantum linear block linear code is a function  $\phi$
from the Hilbert space
of a qubit to a Hilbert space of $m$ qubits:
$\phi: \cal{C}$$^{2} \longmapsto \cal{C}$$^{2^{m}}.$
$m$ is called the size of the block in the code.
Such a code  induces a linear function from $\cal{C}$$^{2^{n}}$
to $\cal{C}$$^{2^{mn}}$ in the following way:
a pure state in $\cal{C}$$^{2^{n}}$, $|\alpha\rangle=\sum_{i} c_i |i\rangle$
will transform to 
$|\beta\rangle=\sum_{i} c_i 
\phi|i_{1}\rangle\phi|i_{2}\rangle...\phi|i_{n}\rangle$.
A pure state in the image of $\phi$ is called a word in the code.
The above definition can be extended to density matrices:
A mixed state of $n$ qubits will be encoded by the corresponding probability 
over the encoding of the pure states.
A mixture of words in the code is also said to be in the code.


The error in the encoded state is sparse if not too many  qubits in each block
are damaged.
Using this notion of sparse sets, we can define a metric 
on block density matrices.
They will be close if the deviation between them
is confined to a sparse set of qubits:

\begin{deff}
Let $B$  be the qubits in $n$ blocks of $m$ qubits. 
A set $A\in B$ of qubits is said to be $k-$sparse
if in each block there are not more than $k$ qubits in $A$.
Two density matrices $\rho_1$ and $\rho_2$ of the qubits $B$
 are said to be $k-$deviated
if there  is an $k-$sparse set of qubits, $A$, such that
  reduced on all qubits except those in $A$ they are equal:
 $\rho_1|_{B-A}=\rho_2|_{B-A}$.
\end{deff}

The deviation is a metric since it satisfies the triangal inequality,
and it is zero only for equal density matrices. 
The quantum code corrects $d$ errors if there is some recovery procedure
$R$, that when applied on all the blocks in any density matrix that
$d$-deviates from a code word $w$, the output is the recovered
 word $w$ (maybe tensored
product with ancilla qubits used by the correction procedure).

\subsection{Quantum computation codes}

A computation code is a quantum code which provides a way to perform
gates on the encoded states fault tolerantly.
The procedure $Pg$ that simulates  a gate $g$ with respect to a quantum code 
is  a sequence of gates which transforms the encoded state to the 
encoded output of the gate: $Pg(\phi(i)\otimes |0>)=\phi(g(i))\otimes |a>$,
where we have used extra ancilla qubits.
These qubits are not counted as the inputs or outputs of the procedure.
A quantum procedure 
is said to have spread $l$ if no qubit or gate effects more than
$l$  outputs of the procedure.
We will need procedures with small spread for fault tolerant computation.
Since we want to convert any arbitrary circuit to a more reliable one,
we need  the set of gates that have $l$-spread
procedures to be universal. 

\begin{deff}
A quantum block code $C$ is said to be a quantum computation code
with spread $l$
if there exists a universal set of quantum gates $G$ such that (1)
for any gate $g\in G$ there exists a
 procedure $P_g$ with respect to $C$,  with spread $ l$,
 and (2) There exist
 correction , initialization and reading  procedures with spread $l$.
\end{deff}

\subsection{Improving reliability by block simulations}

To simulate some circuit by a quantum computation code,
we first convert it to a circuit which uses only gates from 
the universal set of the code. Then we simulate this new circuit 
as was explained:
We now convert each qubit to a block of qubits, each time step to 
a working period,  and each gate
to the corresponding procedure, and besides that we
 add in each working period 
a correction procedure on each block.
Apart from all that, we also add initialization procedures before the first
working period of each computation block and a reading procedure 
after the last working period of each result block.

The space-time of the simulating circuit $M_1$ can be devided to
 rectangles, where each rectangle will correspond to one procedure,
 in the following way:
First, devide the time to alternating stages:
 computation stages, in which one time step of $M_0$ is simulated,
i.e. one level of gate  procedures is applied, and a
corrections stage, in which one level of 
 error correction procedures are applied.
Each stage is a ``strip'' in the space time.
Each strip can be devided to rectangles by deviding the qubits to sets:
A correction strip will be devided such that in each rectangle 
a correction of one qubit is computed.
In a computation strip, we devide the strip to rectangles 
by deviding the qubits to sets, where each set of qubits participates 
in exactly one procedure.
Each rectangle thus corresponds to one procedure.

We show that if a fault path in $M_1$ is such that no more than a few 
faults occured in each rectangle, then indeed the 
computation succeeds. The number of faults allowed in one rectangle 
is bounded so that when taking into account the spread of the fault, 
the number of qubits effected in each block 
at the end of one working period 
is not too big, so that the density matrix can  still be recovered.

\begin{deff}
A fault path of a medium  $M_1$ that block simulates $M_0$
 is said to be a ``$k-$sparse fault path'' 
if no more than $k$ faults occured in each rectangle.
\end{deff}

\begin{lemm}\label{onelevel}
Let $C$ be a quantum computation code that corrects $d$ errors,
with spread $l$.
 Let $M_1$ be a block simulating circuit. 
Consider a computation of $M_1$ subjected to a $k-$sparse fault path
with $d\ge 2kl$.
At the end of each working period
the density matrix is $d-$deviated from the correct one.
\end{lemm}
{\bf Proof:}
We will prove by induction on  $t$ a stronger assertion,
 that at the end of the $t'$th working period
the density matrix is $d/2-$deviated from the correct one.
For $t=0$ the deviation is zero.
Suppose that the density matrix at the end of the $t'$th working period 
 is $d/2$-deviated from the correct 
matrix. After the computation stage, not more than 
$kl$ qubits are effected in each block, so the
density matrix is $kl+d/2$-deviated. Since $kl+d/2\le d$, the correction
procedure indeed corrects the error, only that during
the corrections new errors occur.
Again, the number of effected qubits is not more than $kl$
in each block, and all the other qubits transform as they should, so they 
are corrected. Hence after the correction procedure  the matrix is  
$kl$-deviated. Since  $kl\le d/2$ this proves the induction step.\bbox

We can now compute the effective noise rate of $M_1$.
This will be smaller than $\eta$ if the parameters are chosen correctly:
\begin{theo}
Let $M_1$ simulate  $M_0$ by the computation code $C$,
which corrects $d$ errors,
have spread $l$, with all rectangles smaller than $a$.
 The effective noise rate of $M_1$ is  
  \(\le
 2\left(\begin{array}{c}a\\d/2l+1\end{array}\right)\eta^{\frac{d}{2l}+1}\).
\end{theo}

{\bf Proof:}
If the fault path in $M_1$ is $d/2l$ sparse, the final density matrix
is $d-$deviated from the correct one by lemma \ref{onelevel}.
Measuring all the qubits in the result blocks, and  taking 
majority in each block, gives the correct answer by lemma  \ref{maj}
in the appendixes.
The number of the rectangles in $M_1$ is less than twice the number
of points where faults can occur in $M_0$. 
Therefore the effective noise rate  is smaller than  the probability for 
not more than $d/2l$ faults in two rectangles of $M_1$.
The probability for a rectangle to have more than 
$k$ faults  is smaller
then the number of possibilities to choose $k+1$ points in the rectangle, 
times the probability for these points to have a fault, which gives the result.
\bbox

\section{Concatenated simulations}
%

In this section, we define proper quantum code,
and concatenated simulations by such codes.
We prove that the reliability of the computation
 can be improved to a constant using 
$log(log(n))$  levels of simulations, when the noise  
 is smaller then some constant imposed by the parameters of the code.

\subsection{Improving reliability to a constant}

We define a proper quantum code:
\begin{deff}
A quantum computation code which is associated a set of gates $G$
 is proper if (1) The gate procedures, and the
 initialization, reading
and correction procedures use only gates from $G$, and 
(2) The correction procedure
 takes any density matrix to some word in the code.
\end{deff}

Let $M_0$ be a quantum circuit.
We define recursively  $M_r$,
 an $r$-simulating circuit of a circuit $M_0$ by the proper quantum
computation  code $C$,  as the simulation by $C$ of $M_{r-1}$.
The recursive simulations induce a definition of $s$-blocks:
Every qubit in one stage of the simulation transforms to a block of $m$
 qubits in 
the next stage of the simulation, and this block transforms to $m$ blocks and
so on.
One qubit in $M_{r-s}$ transforms to  $m^s$ qubits in $M_r$.
 This set of qubits in $M_r$ is called an $s$-block.
An $0$-block in $M_r$  is simply a qubit. 
 The recursive simulations also induce a definition of  $s$-rectangles:  
Each space time point in $M_{r-s}$ transforms to a
set of space time points in the following simulation $M_{(r-s+1)}$,
 which in their turn transform to more points in the following
 stages of the simulation.
The set of all these points in $M_r$ that originated from one 
space time point in $M_{(r-s)}$ are called an $s$-rectangle.
The definition of $s$-rectangles defines a devision of the 
space time of $M_r$,  and this devision is 
a refinement of the devision to $(s+1)$-rectangles. 
An $0$-rectangle is just a space time point in $M_r$.

A density matrix of $M_r$ is recovarable if it deviates on a 
 ``sparse'' set of qubits, i.e that in each level, there are enough 
blocks that can be recovered. If at some level there are enough blocks that
can be recovered, the other blocks will be corrected by the error corrections
in the upper level.
\begin{deff}
Let $B$ be the set of qubits in $n$ $r-$blocks .
An $(r,k)$-sparse set of qubits $A$ in $B$  is 
a set of qubits in which for every $r-$block in $B$, there are at most 
$k$ $(r-1)-$blocks such that the set $A$ in these blocks
 is not $(r-1,k)$ sparse. 
An $(0,k)-$sparse set of qubits $A$ is an empty set of qubits.
Two density matrices $\rho_1,\rho_2$, are said to be
$(r,k)$-deviated if there exist an $(r,k)$-sparse set of qubits $A\in B$,
such that $\rho_1|_{B-A}=\rho_2|_{B-A}.$
\end{deff}
The deviation is a metric, as is shown in the appendixes, lemma
\ref{sparsemetric}.
We define ``sparse''  fault paths, that  do not increase the deviation
  in this metric too much: 
\begin{deff}
A set of space time points in an $r-$rectangle 
is said to be $(r,k)$-sparse if  there are no more than $k$ 
$(r-1)-$ rectangles, in which the set is not $(r-1,k)$-sparse.
An $(0,k)$-sparse set in an $0-$rectangles (which is one space time point)
is an empty set. a fault path in $M_r$ is $(r,k)$-sparse
if in each
$r-$rectangle, the set is $(r,k)-$sparse. 
\end{deff}   

We claim  that if the fault path is sparse enough,
then the error corrections keep the deviation small.

\begin{lemm}\label{Theproof}
Let $C$ be a proper code that corrects $d$ errors,
with spread $l$.
 Let $M_r$ be a  medium which $r-$simulates $M_0$ by $C$.      
Consider a computation subjected to an $(r,k)$-sparse fault path
with $2kl\le d$.
At the end of each $r-$working period
the density matrix is $(r,d/2)$-deviated from the correct one.
\end{lemm}
{\bf Proof:}The proof is by induction on $r$.
The structure of the proof goes as follows: we 
first  prove by induction on the number of levels,  $r$,
three assertions together. Assuming that the fault path is $(r,k)$-sparse,
 an $r$-computation stage 
does not cause too much $r-$deviation, 
an $r-$correction corrects a small enough $r-$deviation,
and an $r-$correction brings any state to a density metric which is not 
too $r-$deviated from some word in the code.
The first two assertions applied 
alternately for the $r-$computation and $r-$correction stage in $M_r$,
will give the desired result.
The proof involves many details, and 
 is given in the appendixes.
\bbox


We can now prove the main result of this paper:
\begin{theo}\label{Thetheo}
Let $C$ be a  quantum  computation code,
which corrects $d$ errors,
have spread $l$, and size of all procedures smaller than $a$.
Let $M_0$ be a quantum circuit, with size $s$ and depth $t$.
There exists a quantum circuit 
$M_r$ of size $O(s\cdot polylog(s))$ and depth $O(t\cdot polylog(t))$,
such that in the presence of noise $\eta$ which satisfies 
  \(\left(\begin{array}{c}a\\d/2l+1\end{array}\right)\eta^{d/2l}< 1\)
$M_r$ outputs the correct answer with probability $\ge 2/3$.
\end{theo}
{\bf Proof:}
If the fault path is $(r,d/2l)$-sparse, than the final density matrix is 
indeed $(r,d)$-deviated from the correct one, by lemma \ref{Theproof}.
Measuring all result blocks in a density matrix 
$(r,d)$-deviated from the correct final density matrix, and  taking 
majority in each $r$-block, gives the correct answer by lemma \ref{maj}
in the appendixes.
Hence the  probability for $M_r$
to succeed is larger than the probability for a fault 
path to be $(r,d/2l)$-sparse.  With the assumption on $\eta$,
the probability for one rectangle to have a set of faults which is
 $(r,d/2l)$-sparse can be shown to be  exponentially (in $r$) close to one.
 Again, details can be found in the appendix.
 Taking $r$ to be $O(log(log(V(M_0)))$ makes this probability high enough. 
Since the growth in cost is exponential in $r$, 
(we use codes with a constant size), and the number of levels is
 $ O(log(log(V(M_0)))$, the cost 
is polylogarithmic.
\bbox

{~}

{\bf Remark:} Theorem \ref{Thetheo} requires
that the code can correct $d>1$ errors.
A similar result holds for $d=1$, with the threshold 
  \(\left(\begin{array}{c}b\\2\end{array}\right)\eta< 1\)
where $b$ is the maximal  size of slightly different  rectangles,
defined to  contain   a computation and a correction procedure together.
The proof is almost the same. 
In some cases this threshold is better. \bbox

\section{Explicit proper quantum computation codes}\label{explicit}
Linear quantum codes\cite{CS95} are represented, using classical codes
over $F_p$, and shown to be  proper for $p=2$.
A subclass of linear codes, polynomial quantum codes, is  defined
and shown to be proper.
%
 
\subsection{Linear quantum codes over $F_p$. }
A linear code of length $m$ and dimension $k$ over the field $F_p$
 is a subspace of dimension $k$
in $F_p^m$, where  $F_p^m$ is the $m$ dimensional 
vector space over the field of $p$ elements. 
Given two linear  codes $C_1$ and $C_2$ such that 
 $\{0\}\subset C_2 \subset C_1 \subset F_p^m$ consider
 the following set of quantum  states in the Hilbert 
space $\cal{C}$$^{p^{m}}$:

\[\forall a\in C1: |S_a>=p^{-(m-k)/2}\sum_{v\in C_2}|a+v>.\]
If $(a1-a2)\in C_2$ than $|S_{a1}>=|S_{a2}>$, otherwise
 $<S_{a1}|S_{a2}>=0.$
Hence these states construct a basis for a linear subspace of
the Hilbert space $\cal{C}$$^{p^{m}}$,
with dimention $z=p^{dim(C1)-dim(C2)}$. This subspace is our quantum code.
Define a second basis of this subspace to be:
\[\forall a\in C_2^{\perp}: |C_a>=\frac{1}{\sqrt{z}}
\sum_{b\in C_1/C_2}w^{a\cdot b}|S_b>~ ,~ w=e^{\frac{2\pi i}{p}}.\]

If $C1$ and $C_2^{\perp}$ both have minimum weight $d$,
then the quantum code can correct for $t=\lfloor\frac{d-1}{2}\rfloor$,
by applying classical error corrections with respect to the code
$C_1$, first in the $S-$basis, and than in the $C-$basis.
 The proofs in \cite{CS95} transforms smoothly 
to this general case. 
 

\begin{theo}
For $p=2$, linear codes are proper and have spread $l=1$.
\end{theo}
{\bf Proof:}
The universal set of gates which is associated with the above codes is:

 $|a,b>\longmapsto |a,a+b>$,

 $|a>\longmapsto\frac{1}{\sqrt{2}}\sum_{b}
 (-1)^{ab}|b>$,

 $|a>\longmapsto |1-a>$, 

$|a>\longmapsto (i)^a|a>$, and

  $|a,b,c>\longmapsto |a,b,c+ab>$,

where all the addition and multiplication are in $F_2$(i.e. mod $2$).
This set is universal\cite{Sho96}. 
The correction procedure is done by applying classical error corrections,
with respect to the code $C_1$, transforming to the $C-$basis by
applying bitwise the gate
  $|a>\longmapsto\frac{1}{\sqrt{2}}\sum_{b} (-1)^{ab}|b>$ correcting 
 classically again and rotating back to the $S-$basis.
The initialization is an error correction, with respect to the code $C_2$
and not $C_1$, which corresponds to the quantum error correction to the code 
which consists of  $|S_0>$ alone.
The reading procedure is applied by computing independently $m$ times the 
$a$ from the state $|S_a>$. 
To apply the procedures of all the gates, except the Toffoli gate, we simply 
apply the gate bitwise on all the qubits (if the block size is $m$,
we apply the gate $m$ times).
A detailed description 
 is given in the appendixes.
The spread of all these procedures  is $l=1$.\bbox

\subsection{Polynomial quantum codes}
To correct $d$ errors, set $m=4d+1$
and set $p>m+1$.
Let $\alpha_1,\alpha_2,...,\alpha_m$ be $m$ distinct non zero elements 
of $F_p$ such that the polynomial 
\(G(x)=\Pi_{i=1}^{m}(x-\alpha_i)\) 
has a non-zero  coefficient of  $x^{2d}$. 
(Such $\alpha_i$ exist because $|F_p|>m+1$). Denote by 

\(V_1=\left\{f(x)\in F(x)~|~deg f(x)\le d\right\},\)

\(V_2=\left\{f\in V_1~|~f(0)=0\right\},\)

\(C_1=\left\{(f(\alpha_1),...,f(\alpha_m))~|~f\in V_1\right\}\subset F_p^m,\)

\(C_2=\left\{(f(\alpha_1),...,f(\alpha_m))~|~f\in V_2\right\}\subset C_1.\)

As before, we use the codes $C_1$ and $C_2$ to define the quantum code:
\[\forall a\in F,~~~~~~
|S_a>=\frac{1}{\sqrt|V2|}\sum_{f\in V_1,f(0)=a}
|f(\alpha_1),...,f(\alpha_m)>.\]

\begin{theo}
Polynomial codes are proper quantum computation codes with spread $l=1$.  
\end{theo}
{\bf Proof:}
The universal set of gates used is:

 $\forall~c\in F$, $|a>\longmapsto|a+c>$,

 $|a>|b>\longmapsto |a>|a+b>$,

$0\ne c\in F$: $|a>\longmapsto|ac>$, 

 $|a>|b>|c>\longmapsto |a>|b>|c+ab>$, 

$\forall c\in F$ $|a>\longmapsto w^{ca}|a>$, and

 for $0<r<p$, the Fourier transform:
 $|a>\longmapsto\frac{1}{\sqrt{p}}\sum_{b\in F}w^{rab}|b>.$

Clearly, all classical reversible functions can be spanned by this set.
We find an explicit unitary matrix in the group generated by this set,
which has infinite order.
We than use group representation theory to show  that this group is dense
in $SU(n)$. By \cite{Sol},
the rate of approaching  each finite matrix is exponentially fast.
The initialization,reading and correction procedure are exactly 
as in the general linear code, where transforming between the  
 $S-$basis and the $C-$basis is 
done by the Fourier transform. 
The procedures  $|Sa>\longmapsto|S(a+c)>$, $|Sa>|Sb>\longmapsto |Sa>|S(a+b)>$,
and  $|Sa>\longmapsto|S(ac)>$ 
can be performed by applying pitwise
the corresponding gates.  
Other procedures use interpolation techniques\cite{BGW}.
\bbox

\section{Generalizations and open problems}
The result implies that quantum computation might be practical if the 
noise in the system can be made very small.
The results should 
motivate physicists to achieve lower noise rates, and
theoreticians to develop a theory for proper quantum codes, and seek 
such codes with 
better parameters, to push the threshold as high as possible.
The point at which the physical data meets the theoretical threshold
 is where quantum computation becomes
practical.
 
The results of this paper hold also in the case of 
circuits which allow to operate only on nearest neighbors.
(We thank Richard Cleve for pointing that out to us.)
This is true since the procedures we use, which are of constant size,
can be made, with constant cost, to operate only on nearest neighbors, by
adding gates that swap between qubits.
However, the bound on $\eta$ in this case will be smaller.

We are grateful to I. Cirac and P. Zoller for the following idea:
In most quantum systems, it is reasonable to assume that different 
types of faults occur with different frequencies. 
In such systems, one can improve the bound on the noise rate significantly,
by using in most levels of the simulation a quantum code which can correct
only for the most frequent errors,
while for less frequent errors it is enough to correct only once in
a few levels.
Examples will be given in the final version.
It is therefore important  to have  good understanding of the
 different noise rates for
 different types of faults,
for specific potential physical realizations of quantum computers.


Our scheme requires a polylogarithmic blow-up in the 
depth of the circuit.
Multilinear Quantum codes can reduce
the depth
from a multiplicative factor of $O(log(n)$ to  a factor of 
$O(log(log(n))$, 
but reducing this to a constant, as in the classical case,
 remains an open problem.


\section{Acknowledgments}
We wish to thank Noam Nisan and Peter Shor  
 for helpful discussions and essential remarks.
We wish to thank Thomas Beth for helpful suggestions, and Richard Cleve 
for solving the question of nearest neighbor gates.
We are grateful to Ignassio Cirac and Peter Zoller for the nice idea of 
how to improve the results for specific systems.
\small
\bibliographystyle{plain}
\bibliography{references}
\normalsize

\section{Appendix A}
Here we give the proof that the if the final density 
matrix is $d-$deviated from the correct one, than measuring 
the qubits and taking majority gives the correct answer.

{~}

\begin{lemm}\label{maj}:
Let $\rho 0$ be the density matrix of the mixed state
 $\{\alpha\}=\{p_k,|\alpha_k>\}$, where $|\alpha_k>=\sum_{i}c^k_i|i>$.
Let $\rho 1$ be a density matrix of the mixed state $\{\beta\}=\{p_k,|\beta_k>\}$ 
where is generated from $|\alpha_k>$ by duplicating each qubit $m$ times:
$|\beta_k>=\sum_{i}c^k_i|i_1^mi_2^m...i_n^m>$.
Let $\sigma_1$ be a density matrix which is $d-$deviated from  $\rho 1$,
where $2d<m$. 
Then measuring the qubits in $\sigma 1$, and taking the majority in each 
block generates a distribution on $n-$strings which is equal
to that generated when measuring $\rho 0$. 
\end{lemm}

{\bf Proof:}
This lemma proves that when 
 the fault path is  $k-$sparse,
 at the end of the computation
the density matrix reduced to the result blocks,
 is $d$-deviated from the correct one reduced to the result blocks.
Note that due to the reading procedures,
the correct density matrix is a mixed state in the subspace
 of $\cal{C}$$^{2^{mn} }$,
which is spanned by basic states in which all the coordinates
 in one block are equal, i.e $|i_1^mi_2^m....i_n^m>$.

We will claim that the  distribution $D$ on $n-strings$
generated by measuring all qubits in $\rho 1$ and taking the majority
is the same as the distribution $D'$, generated 
when measuring  $\sigma 1$ and taking the majority.
The probability to get an $n$-string $i$ when measuring  $\rho 0$,
is $\sum_k p_k |c^k_i|^{2}$.
The probability to get an $n$-string $i$ when measuring  $\rho 1$,
and taking the majority is the same.
When measuring the qubits in $A$, in  $\sigma 1$,
one gets the same answer on all the qubits in one block, since
  $\sigma 1$ reduced to $A$ equals $\rho 1$.
Hence, these qubits determine the majority vote, so
the probability to get an $n$-string $i$ when measuring  $\sigma 1$,
and taking the majority is the same as the probability to get this 
$n$-string $i$ duplicated on $A$ when measuring $A$ in  $\sigma 1$,
which is the same as the probability to get this 
$n$-string $i$ duplicated on $A$ when measuring $A$ in  $\rho 1$,
which is the correct probability.\bbox

\section{Appendix B}
We prove that the deviation is a metric:

{~}

\begin{lemm}\label{sparsemetric}:
Let $C$ be the set of qubits in $n$ $r-$blocks.
Let $A,B\in C$ be two  sets of qubits which are $(r,l_1),(r,l_2)$ sparse
respectively.
Then, $A\cap B$ is  $(r,l_1+l_2)$ sparse.
\end{lemm}

{\bf Proof:}
To see this, use induction on $r$.
For $r=0$, the union of two empty sets is empty.
Let us assume for $r$,
and prove for $r+1$.
$A$ is $(r+1,l_1)$ sparse, so there are at most $l_1$ $r-$blocks
which are not  $(r,l_1)$ sparse.
$B$ is $(r+1,l_2)$ sparse, so there are at most $l_2$ $r-$blocks
which are not  $(r,l_2)$ sparse.
In all the other $r-$blocks, the union of $A$ and $B$ is $(r,l_1+l_2)$ sparse
by induction.
So there are at most $l_1+l_2$ $r-$blocks in which the union is not 
$(r,l_1+l_2)$ sparse,
so the union is $(r+1,l_1+l_2)$ sparse.\bbox

{~}

\section{Appendix C}
In this appendix we give the proof of the main theorem, that concatanated
simulation provide noise resistance.
We first prove a lemma and than give the theorem:
{~}

{\bf Lemma \ref{Theproof}:}
Let $C$ be a proper code that corrects $d$ errors,
with spread $l$.
 Let $M_r$ be a   medium which $r-$simulates $M0$ by $C$.      
Consider a computation subjected to an $(r,k)$-sparse fault path
with $2kl\le d$.
At the end of each $r-$working period
the density matrix is $(r,d/2)$ deviated from the correct one.

{~}

{\bf Proof:}We 
first  prove by induction on the number of levels  $r$
 three assertions, together.
\begin{enumerate}
\item\label{comp}
 Consider $n$ $r-$blocks, in a density matrix which is 
$(r,kl)-$deviated from $\phi^r(\rho_0)$,
where $ \rho_0$ is a density matrix of $n$ qubits.
After applying one stage of $r-computations$ on
these blocks, simulating the operation $g$ on  $\rho_0$,
with an $(r,k)$ sparse set of faults,
the density matrix is  $(r,d)$ deviated from $\phi^r(g\circ\rho_0)$.
\item\label{cor} Consider $n$ $r-$blocks, in a density matrix $\rho_r$,
which is $(r,d)$ deviated from a word  $\phi^r(\rho_0)$,
where $ \rho_0$ is a density matrix of $n$ qubits.
 After applying $r-$corrections, on all of the $r-$blocks,
with an $(r,k)$ sparse set of faults,
the density matrix is $(r,kl)$ deviated from $\phi^r(\rho_0)$.
\item\label{totalcor}
 Consider $n$ $r-$blocks, in some density matrix, $\rho_r$.
 After applying $r-$corrections, on all of the $r-$blocks,
with an $(r,k)$ sparse set of faults,
the density matrix is $(r,kl)$ deviated from a word $\phi^r(\rho_0)$,
where $ \rho_0$ is a density matrix of $n$ qubits.
\end{enumerate}

For $r=0$ the proof is trivial. The computations are just faultless,
and $0-$corrections are the identity.
For instructiveness, let us consider also the case of $r=1$.
Claims \ref{comp},\ref{cor}
are satisfied merely because we use a computation code.
Claim \ref{totalcor} is satisfied by the extra restriction 
on the error correction which a proper  code satisfies.
Let us now assume all the claims for $r$, and prove each of the claims 
for $r+1$, starting with claim number \ref{comp}.

\ref{comp}.
We consider one stage of $(r+1)-$computations on
these blocks, $(r+1)-$simulating the operation $g$ on  $\rho_0$.
The stage of computation $(r+1)-$rectangles can be viewed as a sequence
of alternating stages  of  computation $(r)-$rectangles
and correction $(r)-$rectangles.
( The number of stages of $r-$rectangles during this $(r+1)-$ rectangle
 is $w$.) 
Let us consider the density matrices in the trajectory of 
$\phi^{r+1}(\rho_0)$, at the end of  each of these  $r-$stages.
These matrices can be written as $\phi^{r}(\rho^t_1)$.
Where $\rho^0_1=\phi(\rho_0)$
and $\rho^w_1=\phi(g\circ\rho_0)$.
Let us assume for a second
that all the $(r)-$rectangles in the $(r+1)-$stage have $(r,k)$ sparse set of
faults. (This assumption is wrong- in each $(r+1)-$rectangle we might 
have $k$ $(r)-$rectangles in which the faults are not $(r,k)$ sparse,
but we will deal with this in a second.) Let us also assume that the
density matrix we start with is $(r,kl)$-deviated from $\phi^{r+1}(\rho_0)$,
meaning that we do not allow $kl$ $r-$blocks in each $(r+1)-$block to be
more than $kl-$deviated.
With these assumptions,
we now prove by induction on $t$ that applying an $r-$correction stage followed
by an $r-$computation stage, on a matrix which
 is $(r,kl)$-deviated from $\phi^{r}(\rho^t_1)$,
gives a matrix which is $(r,kl)$-deviated from $\phi^{r}(\rho^{t+1}_1)$.
This is true by applying the induction assumption of claims \ref{comp},
and \ref{cor}.
Now, there might still be $k$ $r-$rectangles with more faults than we assumed,
in each $(r+1)-$rectangle,
and in the initial density matrix
 there might still be $kl$ $r-$blocks with more deviations
 than we assumed, in each $(r+1)-$block.
 Since the $(r+1)-$stage simulates a computation procedure
which has a maximal spread $l$, these $r-$rectangles and $r-$blocks 
can effect only $l(kl+l)$ number of $r-$blocks in each $(r+1)-$block,
 at the end of the $(r+1)-$stage- so if $l(kl+l)<d$,
we have that the final density matrix at the end of the $(r+1)-$stage is
$(r+1,d)-$deviated from the correct one.

\ref{cor}.
We consider one stage of $(r+1)-$corrections on
the $n$ $(r+1)-$blocks.
This stage can be viewed as a sequence
of alternating stages  of  computation $(r)-$rectangles
and correction $(r)-$rectangles.
( The number of stages of $r-$rectangles during this $(r+1)-$ rectangle
 is $w$.) 
Let us consider the trajectory of 
$\phi^{r}(\rho_1)$, where $\rho_1$ is a density matrix 
which is $(1,d)-$deviated from some word $\phi(\rho_0)$,
and denote the density matrices in this trajectory 
  at the end of  each  $r-$stages by $\phi^{r}(\rho^t_1)$.
Then since the $(r+1)-$rectangle simulates error correction,
and the simulated matrix is not too deviated, the trajectory which 
starts with   $\rho^0_1=\rho_1$ $r-$simulated,
will end with $\rho^w_1=\phi(\rho_0)$, $r-$simulated.
Now, let us assume, again, 
that all the $(r)-$rectangles in the $(r+1)-$stage have $(r,k)$ sparse set of
faults,
and  that the
density matrix we start with is $(r,d)$-deviated from $\phi^{r}(\rho_1)$.
With these assumptions,
we can now prove by induction on $t$ that applying an $r-$correction stage followed
by an $r-$computation stage, on a matrix which
 is $(r,kl)$-deviated from $\phi^{r}(\rho^t_1)$,
gives a matrix which is $(r,kl)$-deviated from $\phi^{r}(\rho^{t+1}_1)$.
This is true by applying the induction assumption of claims \ref{comp},
and \ref{cor}.
So under the above assumptions, we end up with a matrix which is $(r,d)$-deviated
from $\phi^{r+1}(\rho_0)$.
Now, we actually start the computation 
with a matrix which is $(r+1,d)-$deviated from  $\phi^{r+1}(\rho_0)$.
So most of the $r-$blocks are 
$(r,d)-$deviated from $\phi^{r+1}(\rho_0)$,
except maybe $d$ $r-$blocks in each $(r+1)-$block which are not.
By the induction stage on claims \ref{cor} and \ref{totalcor},
after the first stage of $r-$corrections, 
most of the $r-$blocks are 
$(r,kl)-$deviated from $\phi^{r+1}(\rho_0)$,
except maybe $d$ $r-$blocks in each $(r+1)-$block which are 
$(r,kl)-$deviated from $\phi^{r}(\rho'_1)$.
So after the first $r-$correction the density matrix is 
$(r,kl)-$deviated from $\phi^{r}(\rho_1)$, where $\rho_1$
is $(1,d)-$deviated from $\phi(\rho_0)$.
We can now use the induction from before and say that the final density matrix
is  $(r,kl)$-deviated
from $\phi^{r+1}(\rho_0)$.
We now take into account the fact that there where $r-$rectangles where the 
faults where not $(r,k)-$sparse.
By the fact that the $(r+1)-$correction simulates an $l-$fualt tolerant procedure,
these can effect only $kl$   $r-$blocks in each $(r+1)-$block,
 at the end of the $(r+1)-$stage,
so we have that the final density matrix at the end of the $(r+1)-$stage is
$(r+1,kl)-$deviated from the correct one.

\ref{totalcor}.
We consider one stage of $(r+1)-$corrections on
the $n$ $(r+1)-$blocks. in some density matrix.
This stage can be viewed as a sequence
of alternating stages  of  computation $(r)-$rectangles
and correction $(r)-$rectangles.
( The number of stages of $r-$rectangles during this $(r+1)-$ rectangle
 is $w$.) 
Again, let us assume that the faults in all the  $r-$rectangles
are $(r,k)-$sparse.
By the induction stage on claim \ref{totalcor},
after one stage of $r-$corrections,
the density matrix is 
$(r,kl)-$deviated from some $\phi^{r}(\rho_1)$.
Let us consider the trajectory of $\phi^{r}(\rho_1)$
in the  $(r+1)-$correction rectangle.
Since it is an $r-$simulation of a correction, it takes 
the density matrix to some word $\phi^{r+1}(\rho_0)$.
As before, we can prove by induction on the $r-$stages that 
at the end of the $(r+1)-$rectangle we end up with a matrix which 
is $(r,kl)-$deviated from $\phi^{r+1}(\rho_0)$,
and taking into account the $r-$rectangles with faults which are not 
$(r,k)-$sparse, we end up with a density matrix which is 
$(r+1,kl)-$deviated from $\phi^{r+1}(\rho_0)$.

We can now use claims \ref{cor} and \ref{comp} alternately
to prove by induction on the number of $r-$working period that 
at the end of each $r-$working period in $M_r$ the density matrix
is $(r,kl)-$deviated from the correct one.
\bbox

We can now prove the main result of this paper:

{\bf Theorem \ref{Thetheo}:}
Let $C$ be a  quantum  computation code,
which corrects $d$ errors, have block size $m$, ($2d<m$),
have spread $l$, and size of all procedures smaller than $a$.
Let $M_0$ be a quantum circuit, with size $s$ and depth $d$.
There exists a quantum circuit 
$M_r$ of size $O(s\cdot polylog(s))$ and depth $O(d\cdot polylog(d))$,
such that in the presence of noise $\eta$ which satisfies 
  \(\left(\begin{array}{c}a\\d/2l+1\end{array}\right)\eta^{d/2l}< 1\)
$M_r$ outputs the correct answer with probability $\ge 2/3$.

{~}

{\bf Proof:}
When measuring all qubits in a density matrix 
$(r,d)$-deviated from the correct final density matrix, and  taking 
majority in each $r$-block gives the correct answer by lemma \ref{maj}.
If the fault path is $(r,d/2l)$-sparse, than the final density matrix is 
indeed $(r,d)$-deviated from the correct one, by lemma \ref{Theproof}.
Hence the  probability for $M_r$
to succeed is larger than the probability for a fault 
path to be $(r,d/2l)$-sparse. 
Let us show by induction on $r$ that
the probability of the faults in an $r-$rectangle to be $(r,d/2l)$ sparse
is higher than  \(1-\eta^{(1+\epsilon)^r}\),
where we set \(\left(\begin{array}{c}a\\d/2l+1\end{array}\right)\eta^{d/2l+1}=
\eta^{1+\epsilon}\). We can do that because of the assumption on the 
parameters.
$\epsilon$ is therefore a positive constant.
The probability for an $0-$rectangle, i.e. one space time point,
 to have faults which are $(0,d/2l)$ sparse, i.e. that in this point
a fault did not occur, is $1-\eta$.
For the step of the induction, assume for $r$, and let us prove for $r+1$.
For the faults in an $(r+1)$-rectangle to be $(r+1,k)$ sparse,
there must be at most $k$ $r-$rectangles in which the fault is not 
$(r,k)$ sparse.
So \(P(r+1)\ge 1-\left(\begin{array}{c}a\\d/2l+1\end{array}\right)
(1-Pr)^{d/2l+1}\),
which gives the desired result 
using the induction assumption and the definition of $\eta^{1+\epsilon}$.
This  proves that
 the probability of success of $Mr$, satisfies
\( Pr\ge (1-\eta^{(1+\epsilon)^r})^{2V(M_0)}\) 
 since the number of $r-$rectangles 
is less than $2V(M0)$. Taking $r$ to be $O(log(log(V(M_0)))$ gives $Pr$ 
which is a constant.
Since the growth in cost is exponential in $r$, 
(we use codes with a constant size)
the blowup that is accumulated in $ O(log(log(V(M_0)))$ is polylogarithmic.
\bbox

\section{Appendix D}

This appendix gives a full description of the fault tolerant 
procedures for linear codes.

(1) {\bf Initialization procedure}:

The initialization procedure is actually a correction procedure 
by the code of one word, which is $|s_0>$.
We will need $m^3$ ancilla qubits.
First apply on each qubit in the first block a rotation around the x axis.
After \cite{Sho96}, let us refer to 
 the state $\frac{1}{\sqrt{2}}(|0^m>+|1^m>)$ on $m$ qubits as a ``cat state''.
To generate such a ``cat state'' apply 
 a rotation around the $x$ axis of the first qubit and then a controlled not
from this qubit to all the other $m-1$ qubits.
To compute the syndrome of a word in a quantum code which 
contains only $|s_0>$,  generate $k$ such ``cat states''
on $mk$ ancilla qubits, and add $km$ ancilla qubits which are all in $|0>$.
 To compute the $j'$th bit of the syndrome,
apply  controlled not from
the block we are initializing to each of the ``cat states'' bitwise,
only on the coordinates which in the $j'th$ raw of the 
parity check matrix of $C^{perp}$  are $1$. Now apply a controlled not from
these cat states to the ``blank'' ancilla states, to imitate a measurement
on them. Now compute from each cat state the syndrome bit, using
only gates from the universal set.
To perform the initialization procedure, compute the syndrome
in the above way independently for each qubit, i.e. $m$ times.
>From each $j'$th copy of the syndrome, compute whether the $j'$th 
bit is wrong or not. 
This is done as follows: 
the vector space can be divided to strange cosets of the 
subspace $C^{perp}$. Each coset can be written as $C^{perp}+e$ where $e$ is a
vector. Each such $e$ gives one possible syndrome.(He=s).
Given the syndrome, we compute the {\bf table} $s ->e $,
and decide whether a qubit is wrong by asking whether it is in the support of $e$,
meaning that the corresponding  coordinate in $e$ is $1$. 
Finally apply controlled not from the 
results to the initialized qubits.
After the first rotations we have
 $\sum_{i=0}^{2^m-1}|i>$. The corrections will then take 
this state to an equal distribution over all the basic states in $C^{perp}$.
The fact that this distribution is uniform is due to the linearity of the code.

So the initialization procedure is actually defined by correcting
an ancilla block to $|s_0>$ and than using the block we are interested in 
initializing, and apply a controlled not from each bit to the bit of
the $|s_0>$. This will not change $|s_0>$ if the input string is all 
$0's$ but will convert it to $|S_1>$ if the string is a string of $1's$, 
and the resulting state is the output of the initialization procedure.

The spread of this procedure is $1$: a fault in the cat states can only
effect one qubit at the end. A fault in the first rotations 
and in the first controlled not gates can only effect one bit at the end.
%
%

(2) {\bf Correction procedure:}
For the correction procedure we first create a state $|s_0>$ on ancilla 
qubits.
We than apply 
a correction procedure, exactly as the procedure in the initialization 
procedure, only that instead of computing from the $j'$th copy of a 
syndrome whether the $j'$th bit is wrong, we also compute whether the number
of faults according to the syndrome is larger than $d$.
The controlled not from the result to the $j'th$ qubit is replaced by
a Toffoli gate which also checks if the number of faults is smaller than $d$.
We also add a gate which replaces the qubit with a qubit from the state 
  $|s_0>$ if the number of faults is indeed larger that $d$.
After this, all the qubits are rotated in the $x$ axis and than the 
whole sequence of gates is repeated, and than all the qubits are again 
rotated in the $x$ axis.


(3) {\bf Reading procedure:}
This procedure requires $m(m+1)$ ancilla qubits.
First we apply controlled not from each of the qubits to one of the last $m$ 
ancilla qubits. This simulates $m$ basic  measurements of
 these qubits\cite{AN96}.
We repeat this operation again to each one of the $m$ blocks of $m$ ancilla 
qubits, so we have $m$ copies of the measured state.
On each of these copy we apply an operation that computes the bit
that is represented by this state, using only gates from the universal set.
(This might require more ancilla qubits.)
The resulting qubits are the $m$ qubits which carry the results from
these $m$ computations.
The spread of this procedure is $1$.
A fault in the first stage of copying the qubits $m$ times can only
effect one qubit in each of the copies, and if the number of faults
is smaller than the critical number, the fault has no effect on the resulting
qubits. A fault during the computation of one bit can only effect this bit.
So the spread is $1$, as long as the sum of damage in the block 
 and number of faults in the 
first stage is smaller than half the minimal distance of the code.  

(4) The procedures of a controlled not gate: 
 $|Sa,Sb>\longmapsto |Sa,S(a+b)>$, 
 The Fourier transform, or the rotation around the $x$ axis by $\pi$:
$|Sa>\longmapsto\frac{1}{\sqrt{2}}\sum_{b}
 (-1)^{ab}|Sb>$, the not gate  $|Sa>\longmapsto |S(1-a)>$, 
 and also a rotation around the $z$ axis by $\pi/2$:
  $|Sa>\longmapsto (i)^a|Sa>$, is achieved  on the encoded state
by applying bitwise the gates on each of the qubits
in the block. The procedure's spread is $1$: a fault can only 
effect one qubit.

(5) {\bf Toffoli gate:} we define a Toffoli procedure as is 
described in \cite{Sho96}, where all the measurements are replaced by 
controlled 
not gates. The only piece of this procedure which is not straight 
forward is creating the ancilla state $|A>$.
The way we do that is create a cat state of encoded qubits:
i.e. use $m$ blocks of qubits that we first initialize to the state 
$|S_0>^m$, and rotate the first block, and then copy this block bitwise on all
the other blocks, giving the state:
$\frac{1}{\sqrt{2}}(|S_0>|S_0>...|S_0>+|s_1>|S_1>...|S_1>)$.
create three such ancilla states.
Also create $|A>+|B>$ as in \cite{Sho96}.
Now apply $m$ times the  transformation: 
$|S_a>|b>|c>|d>---> (-1)^{a(bc+d)}|S_a>|b>|c>|d>$
on a block in the first encoded cat state and three bits in the three 
blocks of $|A>$, then do the same thing with the second 
cat state and then the third.
To measure the cat state,
rotate each of the blocks in the cat states in the x direction.
Now compute from each block the bit it represents, and than independently
$m$ times the parity of these bits in each cat state.
For each qubit in the third block of $|A>$,
compare three parity bits from the three cat states 
by a majority vote and apply a controlled not from the result to 
the qubit in $|A>$.
 If only one fault occurred in this procedure, than in each block of $|A>$
there is at most one deviated qubit. This means that the procedure 
is fault tolerant.

\section{Appendix E}
In this section we describe the gate procedures for
the polynomial quantum codes over $F_p$.
The initialization,reading and correction procedure are like for 
general linear codes.

(1)The procedure of a general not gate,  $|Sa>\longmapsto|S(a+c)>$.
A  general controlled not gate: $|Sa>|Sb>\longmapsto |Sa>|S(a+b)>$
and the following operations:
For a fixed constant $0\ne c\in F$: $|Sa>\longmapsto|S(ac)>$
are done by applying the same gates pitwise on the qupits.

(2)The procedure of a general Toffoli gate 
 $|Sa>|Sb>|Sc>\longmapsto |Sa>|Sb>|S(c+ab)>$
is applied as follows:
First we apply pitwise the general Toffoli gate on the $m$ coordinates.
On the third block we obtain the sum:
\[\sum_{A(x),B(x),C(x)\in V1,A(0)=a,
B(0)=b,C(0)=c}
|A(\alpha_1)B(\alpha_1)+C(\alpha_1),...,A(\alpha_m)B(\alpha_m)+C(\alpha_m)>.\]
The polynomial $ D(x)=A(x)B(x)+C(x)$ satisfies $D(0)=ab+c$,
and it's   degree $deg(D)\le 2d$.
To reduce the degree we use a quantum analog of the
 techniques in \cite{BGW}, where we keep the procedure fault tolerant.
we can still correct $d$ errors in $D$ since $m=4d+1$. 
We proceed as follows:
we apply the initialization procedure to each coordinate in the third block,
which will give the state  \[\sum_{D(x), D(x)=A(x)B(x)+C(x)}
|S(d_1),...S(d_m)>~,~ with~ d_j=D(\alpha_j).\]
We then first run the error correcting procedure of the code with degree $2d$,
and after this compute  the linear combination 
\(S(\sum_{l}c_ld_l)=Sd\),
where the  $c_l$ are the interpolation coefficients such that 
\(\forall~ f\in F[x],~ deg(f)\le m-1, f(0)=\sum_{i=1}^{m}c_if(\alpha_i)\).

(3) The procedure 
$|Sa>\longmapsto w^{ca}|Sa>$ $\forall c\in F$, 
(This  generalizes a rotation around the $z$ axis by the angle $\pi$.)
is done by on the $l'th$ qupit
the gate $|a>\longmapsto w^{c_la}|a>$.
The proof that this achieves the desired operation is :
\[|S_a>=
\frac{1}{\sqrt|V2|}\sum_{f\in V,f(0)=a}|f(\alpha_1),...,f(\alpha_m)>
\longmapsto
\sum_{f\in V1,f(0)=a}\Pi_{i=1}^{m}w^{c_lf(\alpha_{l})}
|f(\alpha_1),...,f(\alpha_m)>=\]\[=\sum_{f\in V,f(0)=a}w^a
|f(\alpha_1),...,f(\alpha_m)>.\]

(4) The Fourier transform:
 $|Sa>\longmapsto\frac{1}{\sqrt{p}}\sum_{b\in F}w^{ab}|Sb>.$
 (This can be done with any $w^{r}$ by replacing in the following 
$w$ by $w^r$.
This operation  
generalizes the rotation around the $x$ axis by the angle $\pi$.)

To perform this procedure we first note that
there are fixed non zero \(e_1,...,e_m\) such that 
for any polynomial $f(x)$ over $F_p$ with $deg(f)\le m-1$,
$f_{2d}=\sum_{e_i}e_if(\alpha_i)$.
This is true since interpolation via $\alpha_1,...\alpha_m$ is a linear 
functional.
Denote  $w_l=w^{e_l\alpha_l^{2d}}, l=1,...,m$.
Let us operate on each coordinate, that is qupit, by the Fourier 
transform \[|a>\longmapsto \sum_{b\in F_p}w_l^{ab}|a>.\]
We claim that this indeed gives the desired operation.
This is true since 
\[|Sa>=
\frac{1}{\sqrt|V2|}\sum_{f\in V,f(0)=a}|f(\alpha_1),...,f(\alpha_m)>
\longmapsto\frac{1}{\sqrt{p}}\sum_{b1,b2,..bm\in F}\sum_{f\in V,f(0)=a}
w^{\sum_{l=1}^{m}e_l\alpha_l^{2d}f(\alpha_l)b_l}|b1,..bm>.\]

Let $b(x)$ be the unique interpolation polynomial $b(\alpha_l)=b_l$,
with degree $deg(b)\le m-1$, for some $b_1,...b_m\in F_p$.
We distinguish two cases:

\begin{itemize}
\item
CASE 1: $Deg(b)\le d.$ In this case the polynomial $h(x)=x^{2d}b(x)f(x)$
for $f(x)\in V1$ is of degree $\le 4d=m-1$ and so 
\[\sum_{l=1}^{m}e_l\alpha_l^{2d}f(\alpha_l)b_l=f(0)b(0)=coeff ~of~ x^{2d} ~in~
h(x).\]
In the above sum, we will have:
\[\frac{1}{\sqrt{p}}\sum_{b1,b2,..bm\in F,b(x)\in V1}\sum_{f\in V,f(0)=a}
w^{b(0)f(0)}x|b1,..bm>=\sum_{b\in F_p} w^{ab} |Sb>.\]

\item CASE 2: $Deg(b)> d$.
We claim that the sum vanishes for the $b's$ in this case.
Let $h(x)$ be the interpolating polynomial through the values 
$\alpha_l^{2d}f(\alpha_l)b_l$. Then $h(x)=x^{2d}f(x)b(x)(modG(x)).$
recall the definition og $G(x)$.
The power of $w$ in the sum is the coefficient of $x^d$ in this polynomial.
It is enough to show that this is not always the same value when summing 
over $f\in V1$ with $f(0)=a$, since then the sum vanishes.
Let $r=deg(b)>d$.
Picking $f(x)=a+cx^{2d-r+1}$, $deg(f)\le d$,
and deg $x^{2d}f(x)b(x)=4d+1=m.$
Therefore $h(x)= x^{2d}f(x)b(x)-cB_rG(x)$ where $Br$ is the 
leading coefficient in $b(x)$,
i.e $b(x)=B_rx^r+...+B_0$. $B_r\ne 0$. Looking
at the coefficient of $x^{2d}$ of $h(x)$ we have  
$ aB_0-cB_rg_{2d}$
where $g_{2d}\ne 0$ is the coefficient of $x^{2d}$ in $G(x)$.
This can obtain any value we want by selecting an appropriate $c\in F_p$.

\end{itemize}
\end{document}